\documentclass[a4paper]{statsoc}

\usepackage{fullpage}
\usepackage{natbib}

\usepackage{amscd,amsmath,latexsym,amssymb,amsfonts}
\usepackage{mathrsfs} 

\usepackage{bm}
\usepackage{dsfont}

\usepackage{graphicx}
\usepackage{color}
\usepackage{subcaption}
\usepackage{bm}

\newcommand{\yb}{\bm{y}}

\newcommand{\R}{\mathds{R}}

\usepackage[a4paper,left=2cm,top=2cm,right=2cm,bottom=2cm,ignoreheadfoot]{geometry}

\title{Discussions of the paper \\
``Sparse graphs using exchangeable random measures"\\
by F. Caron and E. B. Fox}
\author{Julyan Arbel}
\address{Inria Grenoble Rh\^one-Alpes, Laboratoire Jean Kuntzmann, France.}

\author{Marco Battiston}
\address{Oxford University, UK.}
\author{Stefano Favaro}
\address{University of Torino and Collegio Carlo Alberto, Italy.}
\email{julyan.arbel@inria.fr, marco.battiston@stats.ox.ac.uk,
stefano.favaro@unito.it,
antonio.lijoi@unibocconi.it, igor@unibocconi.it, 
ramses@sigma.iimas.unam.mx,
yangni87@gmail.com, pmueller@math.utexas.edu}
\author{Antonio Lijoi, Igor Pr\"unster}
\address{Department of Decision Sciences and BIDSA, Bocconi University, Milan, Italy.}
\author{Rams\'es H. Mena}
\address{IIMAS, UNAM, M\'exico.}
\author{Yang Ni, Peter M\"uller}
\address{University of Texas at Austin, Austin, USA.}

\begin{document}
\maketitle 

\begin{abstract}
These are written discussions of the paper ``Sparse graphs using exchangeable random measures"
by Fran\c cois Caron and Emily B. Fox, contributed to the Journal of the Royal Statistical Society Series B.
\end{abstract}

\section{Consistency of the sparsity parameter (J. Arbel)}

The article by Caron and Fox is a very fine methodological work which illustrates once again the huge modeling flexibility and versatility of discrete Bayesian nonparametric priors. They target here sparsity in graphs, the level of which can be neatly assessed according to the stability parameter $\sigma$ of the discrete process under consideration. 

The posterior distribution of $\sigma$ is notoriously highly concentrated in the context of Bayesian nonparametric inference for species sampling problems. The  credible intervals narrowness obtained for the real world graphs suggests that the same hold here. Caron and Fox validate the introduced methodology via posterior predictive checks such as the fit to the empirical degree distribution. Another type of validation, theoretical though, which is not considered by the authors is through \textit{posterior consistency}. In the present setting, the graph is given and interest is in assessing graph properties such as sparsity. Posterior consistency here amounts to ask whether the model is capable of recovering a sparsity index $\sigma$, in other words: 
\textit{If the ``true'' graph generative process is assumed to have a sparsity index $\sigma_0$, 
then does the posterior of $\sigma$ contract to a point mass at $\sigma_0$ when the size of the graph increases to infinity?} 
The sparsity index $\sigma_0$ can be defined   in the spirit of Eqn (21) by the asymptotic relationship between the number of nodes  $N_\alpha$ and the number of edges $N_\alpha^{(e)}$:
\begin{equation*}
\quad\quad\quad\quad\quad\quad\quad\quad\quad\quad\quad\quad	 N_\alpha^{(e)}/N_\alpha^\frac{2}{1+\sigma_0}\to 1 \quad\quad\quad\quad\quad\quad\quad\quad\quad\quad\quad\quad\quad\quad\quad\quad \text{(S)}
\end{equation*}
as the graph size grows to infinity. Accordingly to the definitions given in Eqns (17) and (18), the \textit{graph size} can be equivalently measured by $\alpha$,   $N_\alpha$ or $N_\alpha^{(e)}$.
The true graph generative process could be the Generalized Gamma process itself (well-specified setting) or any other graph process satisfying~(S) for some $\sigma_0$ (mis-specified setting).

In recent research \citep{arbel2017consistency}, we have introduced conditions on the L\'evy intensity of the completely random measure to ensure consistent estimation in  species sampling problems. Though the setting is quite different, our conditions are similar to the tail assumptions made by Caron and Fox in the derivation of $N_\alpha$ and $N_\alpha^{(e)}$ asymptotic behaviors. Admittedly, the consistency assumption of a true generative model with a given fixed sparsity level is an idealized assumption which cannot account for  real world graphs oddities such as local effects underlined by the authors: dense subgraphs (spots), community structure. However we believe that consistency properties could shed some light on why $\sigma$ were estimated to be negative for most of the real world applications in the paper, thus concluding on dense graphs.

\section{Developments related to privacy issues (M. Battiston and S. Favaro)}

We wish to congratulate the authors for the interesting contribution, which has already attracted a lot of interest in the statistical community. Here we would like to point toward new developments related to privacy issues in network modeling. Network data usually contain sensitive information about individuals, e.g. medical status, wages, friendships, sexual or political preferences, etc. A noteworthy example of privacy disclosure is in \cite{Nar09}, where the authors showed how to identify users in the Netflix dataset, which can be modeled by a weighted bipartite graph, even after users and movies labels had been removed. Privacy problems concern with providing mechanisms to transform row data into a privatised dataset to be released. A popular measure to check whether a mechanism is able to privatise a dataset is \textit{differential privacy}, initially proposed in \cite{Dwo06} and recently considered in graph theory for network data. A mechanism $\mathcal{A}$  is said to be \textit{node private} if an intruder looking at the output released by the mechanism cannot correctly guess with high probability if a node (individual) is in the dataset or is not and figure out which are his edges (links). A formal definition of \textit{$\epsilon$-node privacy} is that for all subset $S$ of the output space, the mechanism $\mathcal{A}$ must satisfy 
 \begin{equation} \label{privacy}
 \text{Pr}(\mathcal{A}(G)\in S)\leq \exp (\epsilon)\text{Pr} (\mathcal{A}(G')\in S)
\end{equation}  
for all graphs $G,G'$ which can be obtained one from the other by removing a vertex and its adjacent edges.

Up to our knowledge, the only attempt to study an \textit{$\epsilon$-node private} mechanism is in \cite{Cha15}. They consider sparse graphs obtained by rescaling a dense graphon with the network size, and they propose a mechanism that releases as output a step graphon that satisfies \eqref{privacy}. We believe that an interesting line of research would be to study how privacy constraints could be handled within the  sparse graphs setting proposed by Caron and Fox. Specifically, is \textit{$\epsilon$-node privacy} a good measure of disclosure for graphs or better notions could be needed in the sparse regime? How to construct mechanisms satisfying these privacy notions? Will the privatized network dataset obtained by this mechanism preserve enough statistical utility? As pointed out in \cite{Nar09}, sparsity facilitates disclosure and at the same time it makes statistical inference more difficult. Therefore, on the one hand, we might need a quite stringent notion of privacy for sparse graphs, but on the other hand this requirement may drastically affect the statistical utility of the released network dataset. As a consequence, a clear trade-off between privacy guarantees and statistical utility arise, particularly in sparse settings. Natural questions are how to mathematically formalize this trade-off and then how to solve it.

\vspace{-.3cm}

\section{Extension to a multi-sample context (A. Lijoi, R.H. Mena, I. Pr\"unster)}

We congratulate the authors for proposing a clever construction of random graphs for networks, which allows to achieve sparsity in an effective way. We are pleased to see another instance of successful use of completely random measures (CRMs) for building flexible Bayesian nonparametric procedures.
 
Among several potential developments, of particular interest is the extension to a multi--sample context with data recorded from two or more networks.  
Assuming the data to be still conditionally independent but not identically distributed, a natural problem 
is 
the derivation of   
testing procedures to verify whether the probability distributions, or some of their features, are shared across different samples.  
For instance, for the GGP--based model the  parameter $\sigma$ plays a pivotal role: as shown in the paper it determines the sparsity properties of the graph, but it also influences posterior inferences on the clustering of the data (see, e.g., \citealp{jrssb_2007}).  Given this, one may assess similarity of two networks directed by two independent GGPs by performing a test on the equality of their respective $\sigma$ parameters: if the test concludes they coincide then the two networks are deemed to be homogeneous. Along these lines, in \cite{jcb} a Bayesian test on the discount parameter of the Pitman--Yor process is defined in order to compare samples of expressed sequence tags from tissues generated under different experimental conditions. Note that the discount parameter of the Pitman--Yor process plays the same role as $\sigma$ for the GGP.

Alternatively, networks' comparison can be faced by assuming a richer model 
 with dependent CRMs $(W_{1,\alpha},W_{2,\alpha})$  
 accommodating for a wide spectrum of dependence structures across networks, ranging from exchangeability   
 to unconditional independence. In this framework, there is no need to focus on a specific feature of the distribution and one may test whether the two distributions themselves are equal. Recently, unrelated to networks' applications, \cite{bdh:2015} address the issue within a parametric model and provide an insightful discussion on the notion of approximate exchangeability. 
 A natural model that serves the purpose is based on ideas in \cite{ndp:2008}, where the nested Dirichlet process is introduced  for clustering probability curves. Similarly, this approach may be useful for clustering networks based on distributional similarity. While \cite{ndp:2008} consider random probability measures, here one needs a model able to handle CRMs. A potentially fruitful approach is proposed in \cite{camerlenghi}. Let 
 $\tilde q$ be a discrete random probability measure on the space of boundedly finite measures $\mathds{M}_{\R_+}$ on $\R_+$, while $q_0$ is the probability distribution of a CRM on $\R_+$. If $(W_1^*,W_2^*,W_0^*)|\tilde q \sim\tilde q^2\times q_0$, define $(W_1,W_2)=(W_1^*+W_0^*,\, W_2^*+W_0^*)$. Discreteness of $\tilde q$ implies that with positive probability $W_1^*=W_2^*$, which in turn yields $W_1=W_2$. This corresponds to similarity of the networks as they have the same distribution.

\section{Inferring the latent network structure (Y. Ni and P. M\"uller)}

We congratulate the authors on a very interesting paper.
Our discussion highlights a particular use of the proposed models
that we felt was missing in the paper.
Implicit in the paper is an assumption that (part of) the random graph
$D$ (or a derived undirected graph $Z$) is observable. While this is 
common for social network data, 
it is less common in biomedical inference where the
goal is often to infer an unknown latent network structure. 

The typical inference is set up under a hierarchical model
\begin{equation*}
  \label{hie}
  \yb_i\sim p(\yb_i \mid \beta),~~
  \beta \sim p(\beta \mid D),~~
  p(D \mid \phi),~~
  \phi\sim p(\phi),
\end{equation*}
where $p(\beta \mid D)$ maps the graph to the parameters $\beta$ 
(Caron and Fox already used up all other Greek letters) of the
top-level sampling model for the observed data $\yb$. This could be, for example, a Gaussian
graphical model for protein activation $\yb_i$.
And we discuss another example below.
We suggest to use the proposed novel models for $p(D \mid \phi)$.
Good prior regularization is more important in this context than 
in applications where the network is observed.

We illustrate our suggestion with a small simulation study and an
application. 
Both are based on directed cyclical graphs (DCG)
\citep{ni2016reciprocal}, a special case of reciprocal graphical models (RGM)
\citep{koster:96}.  
The DCG allows inference on a directed graph $G$,
possibly including cycles, by setting up a simultaneous equation model
and interpreting a directed edge $(\ell, i)$ in
the graph $G$ as an indicator for a non-zero coefficient of $y_\ell$
in the equation for $y_i$.
In this context we explore the use of a GGP prior $p(D \mid \phi)$,
including a mapping of a multigraph $D$ to a directed graph $G$ by mapping
$n_{ij} \mapsto I(n_{ij}>0)$.

Table 1 reports summaries for a simulation study with
four alternative priors $p(D \mid \phi)$. GGP wins.
Figure 1 shows the estimated graphs in inference
for a gene network reported in \cite{ni2016reciprocal} under 
the sparsity prior used there  (thresholded prior) versus
the new GGP prior.
Also under the realistic conditions of this data analysis the 
choice of prior matters. 
Importantly, implementation of posterior inference was quite straightforward, as
described in the paper. 

\begin{center}
\begin{minipage}{7cm}
\textbf{Table 1.} Simulation study. True positive rate (TPR) and false
    discovery rate (FDR) under 
    four prior models:  Erd\H{o}s-R\'enyi graph with
    $p=0.5$ (ER); ER with $p \sim Be(0.5,0.5)$ (ER+Beta); GGP; and 
    thresholded prior (TP).\bigskip
    
	\begin{tabular}{l|cccc}
		&ER$(1/2)$&ER+Beta&GGP&TP\\
\hline
		TPR&1.00&1.00&0.87&0.87\\
		FDR&0.58&0.72&0.07&0.18
	\end{tabular}
\end{minipage}	
\end{center}

\begin{figure}
\centering
\begin{tabular}{cc}
  \includegraphics[width=0.4\textwidth]{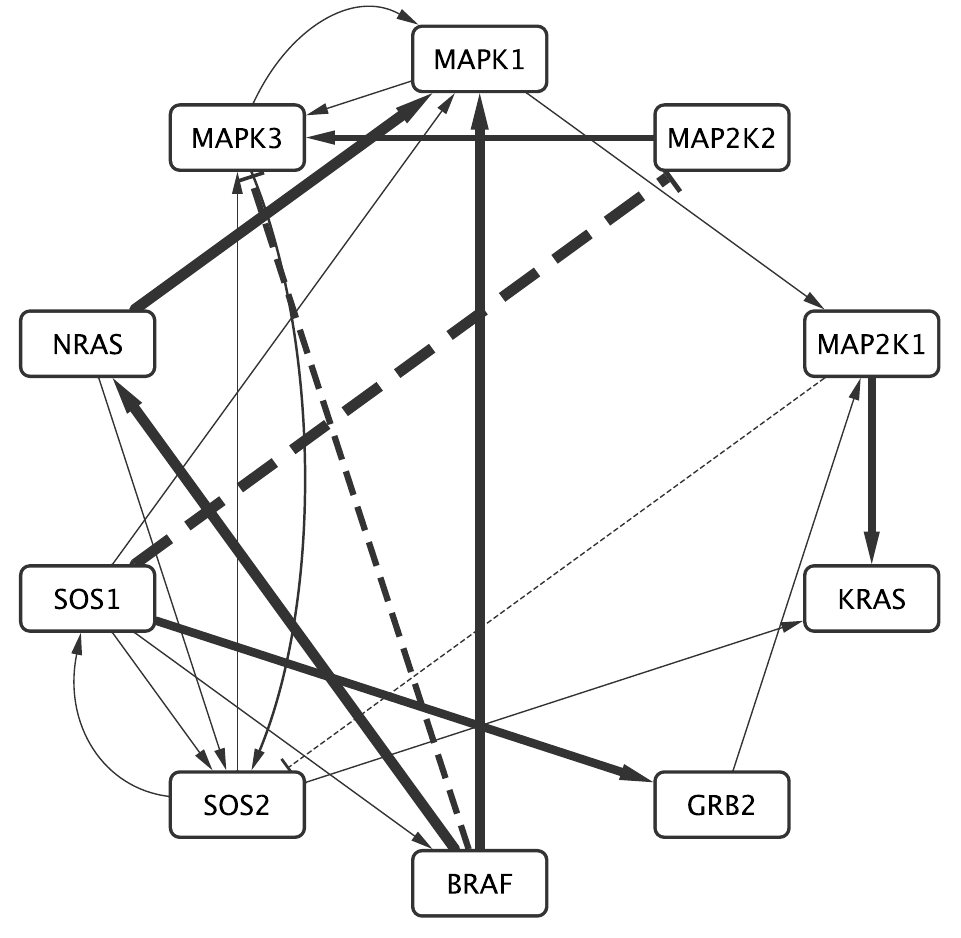} &
  \includegraphics[width=0.4\textwidth]{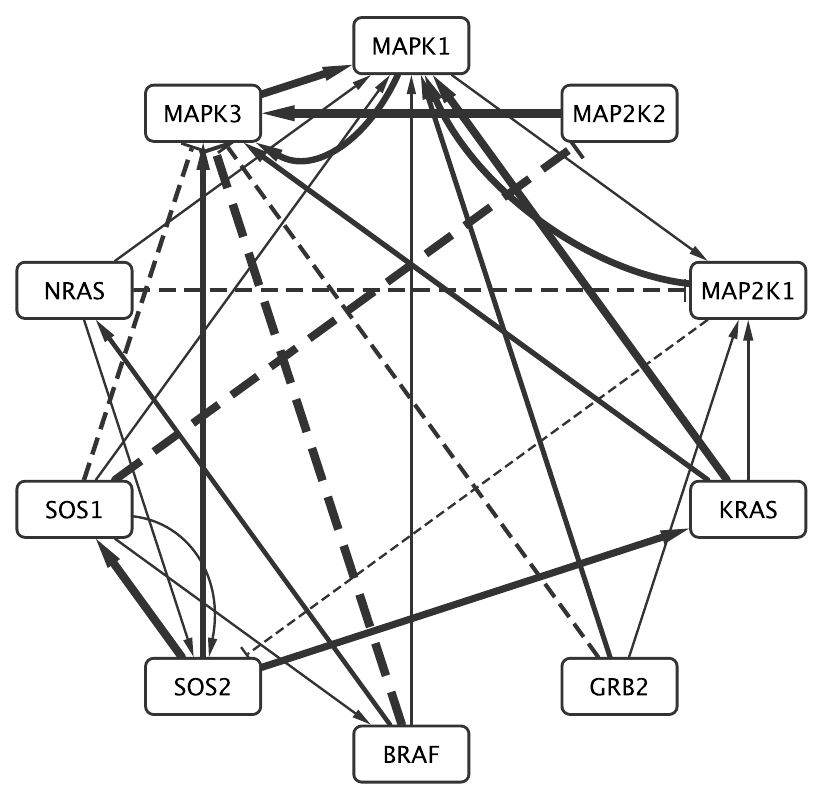}
\end{tabular}
\caption{Estimated gene networks for TCGA colon cancer data under the GGP prior
  (left) and the thresholded prior \citep{ni2016reciprocal} (right).
  Posterior expected FDR is controlled to be less than 10\% for both
  estimations. }
\end{figure}

\end{document}